\documentstyle[12pt,aps,prd,epsfig]{revtex}

\newcommand{\M}{{\cal M}}
\newcommand{\eps}{\varepsilon}

\newcommand{\dels}{\delta q^s}
\newcommand{\delv}{\delta q^v}

\newcommand{\ai}{a_{1}}
\newcommand{\bi}{b_{1}}
\newcommand{\hi}{h_{1}}
\newcommand{\gmnn}{g_{\mbox{$\scriptscriptstyle {\M N N}$}}}

\newcommand{\be}{\begin{eqnarray}}
\newcommand{\ba}{\begin{array}}
\newcommand{\ea}{\end{array}}
\newcommand{\ee}{\end{eqnarray}}

\newfont{\fib}{cmfi10 at 10pt}

\newcommand{\A}{{\cal A}}

\begin{document}

\draft
\preprint{}

\title{\hfill{\rm TUHEP-TH01-04}\\
Estimates of the Nucleon Tensor Charge}
\author{Leonard Gamberg \thanks{Electronic address: gamberg@dept.
physics.upenn.edu}}
\address {Department of Physics and Astronomy, University of
Pennsylvania\\
and\\
Department of Physics and Astronomy, Tufts University}
\author{Gary R. Goldstein  \thanks{Electronic address:ggoldste@emerald.
tufts.edu}}
\address {Department of Physics and Astronomy, Tufts University}

\date{June 15, 2001}
\maketitle

\begin{abstract}
\scriptsize
Like the axial vector charges, defined from the forward nucleon matrix
element of the axial vector current on the light cone, the nucleon tensor
charge, defined from the corresponding matrix element of the
tensor current, is essential for characterizing the momentum and spin
structure of the nucleon. These charges, which are first moments of the
quark helicity and transversity distribution functions, can be measured,
in principle, in hard scattering processes. Because there must be a
helicity flip of the struck quark in order to probe the transverse spin
polarization of the nucleon, the transversity distribution (and thus the
tensor charge) decouples at leading twist in deep inelastic scattering,
although no such suppression appears in Drell-Yan processes. This makes
the tensor charge difficult to measure and its non-conservation makes it
difficult to predict. While there may be constraints on the leading twist
quark distributions through positivity bounds (e.g. the inequality of
Soffer), there are no definitive theoretical predictions for the tensor
charge, aside from model dependent calculations (e.g. the QCD sum rule
approacch).  We pursue a different route. Exploiting the
approximate mass degeneracy of the light
axial vector mesons ($a_1$(1260), $b_1$(1235) and $h_1$(1170)) and using
pole dominance, we calculate the tensor charge. The result is
simple in form and depends on the decay constants of the
axial vector mesons and their couplings to the nucleons,
along with the average transverse momentum of the quarks in the
nucleon. The result is compared with other model estimates.
\end{abstract}

\section{Introduction}

The spin composition of the nucleon has been intensely studied. Its study
has produced important and surprising insights, beginning with the
revelation that the longitudinal spin is carried more by gluons than
quarks. Considerable effort has gone into understanding, predicting and
measuring the corresponding transversity distribution of the nucleon
constituents. The leading twist transversity structure function, $h_1(x)$,
is as fundamental to understanding the nature of the non-perturbative QCD
regime of hadronic physics as is the longitudinal function $g_1(x)$, yet
the transversity distribution can not be measured directly in deep
inelastic lepton scattering since it is chiral odd.

Like the isoscalar and isovector axial vector charges
defined from the forward nucleon matrix element of the
axial vector current,
the nucleon tensor charge is defined from the corresponding
forward matrix element of the tensor current,
\begin{equation}
\langle P,S_{T}\left|\overline{\psi}
\sigma^{\mu\nu}\gamma_5 \lambda^a\psi\right| P,S_{T}\rangle
=2\delta
q^a\left(\mu_0^2\right)\left(P^{\mu}S^{\nu}_T-P^{\nu}S^{\mu}_T\right),
\label{eq1}
\end{equation}
${S_T}^Q \sim \left(\, |+> \pm |->\, \right)$ for the moving nucleon is the {\it
transversity}, a variable introduced originally by Moravcsik and
Goldstein~\cite{transversity} to reveal an underlying simplicity in
nucleon--nucleon spin dependent scattering amplitudes.
Unlike the axial vector isovector and isosinglet charges, no sum rule
has been written that enables a clear relation between the tensor charge
and a low energy measurable quantity. So, aside from model calculations,
there are no definitive theoretical predictions
of the tensor charge. Among the various approaches, from the QCD sum rule
\cite{jihe,jintang,bely},
which estimates the flavor contributions to the tensor charge by
analyzing the bilocal tensor current on the light cone,
to lattice calculations~\cite{aoki} and light cone quark
models~\cite{melosh}, there appears to be a range of expectations and a
disagreement concerning the sign of the down quark contribution.

Given the numerous experiments at, RHIC-Brookhaven
(the PHENIX and STAR collaborations~\cite{phen,star}),
CERN (COMPASS experiment~\cite{comp}),
and HERA-DESY (the HERMES experiment~\cite{her})
and  proposals~\cite{teslsa,ELFE}
for extracting quark transversity distributions and, in turn,
the flavor contributions to the nucleon's
chiral odd tensor charge,  there is reason to expect
that in the not too distant future one will be able
to reliably compare the data to the  theoretical estimates of these
quantities.

The various charges, which are first moments of the quark helicity and
transversity
distribution functions $\Delta q^{a}(x)$ and $\delta q^{a}(x)$
respectively, in principle can be measured in hard scattering processes.
In their systematic study of the chiral odd distributions,
Jaffe and Ji~\cite{jaffe91} related the first moment of the transversity
distribution to the flavor contributions
to the nucleon tensor charge:
\be
\int_0^1 \left(\delta q^a(x)-\delta\overline{q}^a(x)\right) dx=\delta q^a .
\ee
Because there must be a helicty flip of the struck quark
in order to probe the transverse spin polarization
of the nucleon,  the transversity distribution
(and thus the tensor charge) decouple  at leading twist in deep
inelastic scattering.
No such supression appears in Drell-Yan scattering where
Ralston and Soper~\cite{ralston79}
first encountered the transversity distribution
entering  the corresponding  transverse spin (both the
beam's and target's spin being transversly polarized to
the incident beam direction)
asymmetries~.\footnote{The quark lines are not correlated in the hard
quark target amplitude.}   Consequently, the charge is difficult
to measure and its non-conservation~\cite{artru}
makes it difficult to predict.
While bounds placed on the leading twist quark distributions
through positivity constraints suggest
that they  satisfy the inequality of Soffer~\cite{soffer95};
\be
\left|2\delta q^a\right|\le q^a +\Delta q^a ,
\ee
(where $q^a$ denotes the unpolarized quark distribution),
model calculations yield a range of theoretical predictions~\cite{review}.

Here we offer another estimate.
Our motivation stems in part from the result
that the tensor charge does not mix with gluons under QCD
evolution and therefore behaves as a nonsinglet
matrix element.~\footnote{This is to be contrasted with the axial charges.}
This, in conjuction with the fact that the
tensor current is charge conjugation odd (it does not mix
quark-antiquark excitations of the vacuum, since the latter is charge
conjugation even, nor does it mix with
gluonic operators under evolution, since gluonic operators are even),
suggests that the tensor charge is more amenable to a  valence quark
model analysis. With this in mind  we estimate the
tensor charge by using axial vector dominance
and an approximate phenomenological mass symmetry among the
light axial vector mesons ($a_1(1260), b_1(1235)$ and $h_1(1170)$).
The $b_1$ and $h_1$ could couple to the quark tensor current in the
nucleon at low energies, and via the symmetry, their coupling to the
leptons is  related to the $a_1$ production in $\tau$ decay.

Next we present our determination of the tensor charge
under conditions of axial vector
dominance in the context of our phenomenological
symmetry represented by $SU(6)_W \otimes O(3)$ spin-flavor symmetry.
We use this symmetry  to
relate the parameters of the previous section to measurable quantities.
Then we present our results for the isoscalar and isovector and, in turn,
up and down quark contributions to the the tensor charge of the nucleon.
Finally we compare our results  with 
QCD sum rule and other models calculations.

\section{Tensor Charge}

We now apply axial vector dominance to estimate the tensor charge.
That is, we choose to determine the value of the tensor charge at a scale
where the matrix element of the tensor current is dominated by the lowest
lying axial vector mesons~\cite{wein67}.
Under these conditions, the  matrix element of the tensor
current, Eq.~(\ref{eq1}) becomes
\be
\sum_{\M} \frac{\langle 0\left|
\overline{\psi}
\sigma^{\mu\nu}\gamma_5 \frac{\lambda^a}{2}
\psi
\right|\M\rangle\langle \M , P,S_{T}| P,S_{T}
\rangle}{M^2_{\M}-k^2} .
\ee
The summation is over those mesons with quantum numbers,
$J^{PC}=1^{+-}$ that  couple to the nucleon via the tensor current;
namely  the charge conjugation odd axial vector mesons -- the isoscalar
$h_1(1170)$ and the isovector $\bi(1235)$.
To analyze this expression in the limit $k^2\rightarrow 0$
we require the vertex functions for the nucleon coupling to the
$h_1$ and $\bi$ meson and the corresponding matrix elements
of the meson decay amplitudes which are related to the meson to vacuum
matrix element via the quark tensor current.
The former yield the nucleon coupling constants
$\gmnn$ defined from the matrix element
\be
&&\langle M P| P\rangle=
\frac{i\gmnn}{2M_N}\overline{u}\left(P,S_{T}\right)
\sigma^{\mu\nu}\gamma_5 
u\left(P,S_{T}\right)\varepsilon_\mu P_\nu ,
\ee
where $P_\mu$ is the nucleon  momentum,
and the latter yield the meson decay constant, $f_{\M}$
\be
\langle 0\left|
\overline{\psi}
\sigma^{\mu\nu}\gamma_5 \frac{\lambda^a}{2}\psi
\right|\M\rangle=
if^a_{\M}\left(\epsilon_\mu k_\nu-\epsilon_\nu k_\mu\right) ,
\ee
where the $k_\mu$ and $\eps_\nu$ are the meson momentum and
polarization respectively.
For transverse polarized Dirac particles, $S^\mu=(0,S_T)$
we project out the tensor charge using the
constraint on the vector meson, $k\cdot\eps_{\scriptstyle \M}=0$
\be
\delta q^a(\mu^2) =
\frac{ f^a_{\M}\  \gmnn \left(S_T\cdot k\right)\A }{P\cdot\eps \
M_{\M}^2} ,
\ee
where
\be
\A=\frac{ P \cdot\eps\left(S_T\cdot k\right)}{2\ M_N}
\ee
is the nucleon-meson vertex function.
In order to evaluate the tensor charge at the scale dictated
by the axial vector meson dominance  model we must determine the
isoscalar and isovector meson coupling constants.
We take a hint from the valence interpretation of the tensor
charge and  exploit the phenomenological mass
symmetry among the lowest lying axial vector mesons that couple to the
tensor charge; we adopt the spin-flavor symmetry characterized by an
$SU(6)_W \otimes O(3)$~\cite{sakita,close} mulitplet structure.
Thus, the $1^{+-}$ $h_1$ and $b_1$ mesons fall into a
$\left(35\otimes L=1\right)$ muliplet that contains
$J^{PC}=1^{+-},0^{++},1^{++},2^{++}$ states.
This analysis enables us to relate the $a_1$ meson
decay constant measured in $\tau^-\rightarrow a_1^- + \nu_\tau$
decay~\cite{birkel,tsai}
\be
f_{\ai} =(0.19\pm 0.03) {\rm GeV^2} ,
\ee
and the $a_1 N N$ coupling constant
\be
g_{\ai NN}=9.3\pm 1.0 ,
\ee
to the meson decay constants, $f_{\bi} $ , $f_{\hi} $ and coupling
constants, $g_{\bi NN}$ and $g_{\hi NN}$.
We find
\be
f_{\bi}=\frac{\sqrt{2}}{m_{\bi}}f_{\ai}    \, , \quad
g_{\bi NN}=\frac{5}{3 \sqrt{2}} g_{\ai NN} ,
\ee
where the 5/3 appears from the $SU(6)$ factor (1+F/D) and the $\sqrt{2}$
arises from the L=1 relation between the $1^{++}$ and $1^{+-}$
states.~\footnote{We note that our value of $f_{\bi}\approx 0.21\pm 0.03$
agrees well with an independent sum rule determination of $0.18\pm 0.03$\protect\cite{ball,bely}}
These, in turn, enable us to determine the isovector and isoscalar
contributions
\be
\delv =f_{\bi}g_{\bi NN} \frac{\langle k_{\perp}^2\rangle}{\sqrt{2} M_N
M_{\bi}^2} , \quad{\rm and }\quad
\dels =f_{\hi}g_{\hi NN} \frac{\langle k_{\perp}^2\rangle}{\sqrt{2} M_N
M_{\hi}^2}
\ee
respectively (where, $\delv =\left(\delta u - \delta d\right)$, and 
$\dels =\left(\delta u + \delta d\right)$).  
Transverse momentum appears in these expressions because the tensor
couplings involve helicity flips that carry kinematic factors of
momentum transfer. The intrinsic $k_{\perp}$ of the quarks in the
nucleon is non-zero, as determined from Drell-Yan and heavy vector boson
production processes. Using a Gaussian momentum distribution, and
letting $\langle k_{\perp}^2\rangle$ range from
$\left(0.58 \:{\rm to}\: 1.0  {\rm GeV}^2\right)$~\cite{ellis}
results in the up and down quark transversity ranging from
\be
\delta u(\mu_0^2)=(0.53 \:{\rm to}\: 0.92)\pm 0.20   \quad \delta
d(\mu_0^2)=-(0.33 \:{\rm to}\: 0.58)\pm 0.20.
\ee

This range of values for the $u$-quark tensor charge lies slightly lower
than most other estimates while the $d$-quark charge is negative and of a
comparable magnitude. Note that many predictions have the
ratio $\delta d/\delta u$ near -1/4 or $(1-\sqrt{3})/(1+\sqrt{3})$,
the value resulting
from an $SU(3)$ degeneracy between the $\pi^0$ and the $\eta(8)$ octet
elements in their coupling to the $u$-quark and the $d$-quark, i.e. the
isoscalar coupling to $u$ and $d$-quarks is $1\sqrt{3}$ times the
isovector. In our calculation the isoscalar and isovector axial vector
couplings to the nucleon also enter as factors in the expressions for
the charges, with the D/F ratio being 3/2 in exact $SU(6)$. Loosening the
$SU(6)$ constraint and incorporating mixing of the $h_1(1170)$ with the
$h_1(1380)$ will alter the $u$ to $d$ ratio. These variations are being
explored. Several other model calculations are summarized in the
Table~\ref{tab_1}.
\parbox[t]{16cm}{
\begin{table}[t]
\begin{center}
\begin{tabular}{|c||cccccc|}
$\delta q^a$  & Lat & ~SR & ~BAG & ~CQ & ~QS & NJL  \\
\hline
$\delta q^u_T(Q^2)$ & 0.84  & 1.33 & 1.09  & 1.19 & 1.07 & 0.82 \\
\hline
$\delta q^d_T(Q^2)$ & -0.23  & 0.04 & -0.27 & 0.12  & -0.38 &  -0.07\\
\end{tabular}
\vspace{1cm}
\caption{
Sampling of estimates of
the flavor contribution to the nucleon tensor charge
calculated in: lattice (Lat)
\protect\cite{aoki} ;
QCD sum rules (SR)
\protect\cite{jihe} ; the bag model (BAG)
\protect\cite{jaf,jihe} ;
the constituent quark model with
Goldstone boson effects (CQ)
\protect\cite{suz} ;
a quark soliton model
calculation (QS)
\protect\cite{kim_boch} ;
and the NJL chiral soliton
model (NJL)
\protect\cite{rein,gamweig}
 with the associated momentum ranging from $0.40-1.0$
${\rm GeV}^2$ (errors range from 10-40$\%$).}
\label{tab_1}
\end{center}
\end{table}}
The calculation has been carried out at the scale $\mu_0\approx 1$ GeV,
which is set by the nucleon mass as well as being the mean mass of the
axial vector meson multiplet. The appropriate evolution to higher scales
(wherein the Drell-Yan processes are studied) is determined by the
anomalous dimensions of the tensor charge~\cite{artru} via
\be
\delta q(\mu^2)= \left(\frac{\alpha(\mu^2)}{\alpha(\mu_0^2)}\right)
^{\frac{4}{33-2n_f}}\delta q(\mu_0^2).
\ee
This is straightforward but slowly varying.

In conclusion, our axial vector dominance model with
$SU(6)_W \otimes O(3)$ coupling relations provide simple formulae for the
tensor charges. This simplicity belies the considerable subtlety of the
(non-perturbative) hadronic physics that is summarized in those
formulae. We obtain the same order of magnitude as most other
calculation schemes. This result supports the view that the underlying
hadronic physics, while quite difficult to formulate from first
principles, is essentially a $1^{+-}$ meson exchange process. Forthcoming
experiments will begin to test this notion.

\section*{Acknowledgements}

G.R.G. thanks X. Ji for a useful discussion and R. L. Jaffe for bringing
the tensor charge problem to his attention.  This work is supported in 
part by a grant from the U.S. Department of Energy  \# DE-FG02-92ER40702.

\end{document}